\documentclass[%
 reprint,
 amsmath,amssymb,
 aps,
]{revtex4-2}

\usepackage{graphicx}% Include figure files
\usepackage{dcolumn}% Align table columns on decimal point
\usepackage{bm}% bold math
\usepackage{amsmath}
\usepackage{slashed}
\usepackage{revsymb}
\begin{document}

\preprint{APS/123-QED}

\title{Electron Wave Spin in Excited States}% Force line breaks with \\
%\thanks{Leijing}

\author{Ju Gao and Fang Shen}
\email{jugao2007@gmail.com}
\affiliation{University of Illinois, Department of Electrical and Computer Engineering, Urbana, 61801, USA}



%\affiliation{}%
%\affiliation{
%\affiliation{%

% \collaboration{CLEO Collaboration}%\noaffiliation

\date{\today}% It is always \today, today,

\begin{abstract}
The wave spin of an electron can be fully characterized by the current density calculated from the exact four-spinor solution of the Dirac equation. In the excited states of the electron in a magnetic field-free quantum well, the current density has a multiple vortex topology. The interaction of the current with a magnetic potential produces a finer structure of anomalous Zeeman splitting. When the magnetic potential is comparable to the size of the individual vortices, fractional or zero spin effects can be observed.
%\begin{description} for quantum computing.
%\end{description}
\end{abstract}


%\keywords{Spin, quantum computing, Dirac, Maxwell, Shrodinger}%Use showkeys class option if keyword
\maketitle

%\tableofcontents

\section{\label{sec:Spin}Electron wave spin described by current density}
Spin is a fundamental property of an electron that represents the electron's internal angular momentum. However, what actually spins remains an open question, because the particle spin interpretation would require the electron to spin on its own axis at a speed greater than that of light. In electromagnetism, the charge behavior is fully described by the Lorentz covariant four-current, defined as $j=(c \rho, \pmb{j})$, where $\rho$ and $\pmb{j}$ stand for the charge and current densities, respectively, since it is responsible for both the generation and interaction of an electromagnetic field. This leads to the question: could the spin also be described by the current?

In a recent paper~\cite{GaoJOPCO22}, we have shown that a stable circulating current density exists for a Dirac electron in a quantum well without a magnetic field. The circulating current density, denoted by $j$, forms a spinning vortex around the center of the charge density, often referred to as an electron cloud. Expressing the current density in terms of the charge density and a spinning velocity distribution $v(\pmb{x})$ in $j(\pmb{x})=\rho(\pmb{x}) v(\pmb{x})$, we find that $v(\pmb{x})$ is limited to the speed of light everywhere in space. In other words, the entire electron wave, or the electron cloud, spins.

Essentially, this is the wave spin interpretation that was first proposed by Belinfante~\cite{Belinfante39,Belinfante40} who argued that spin should be regarded as a circulating flow of energy of the electron field. Ohanian~\cite{Ohanian86} further elaborated the connection between the circulating momentum density and current density with the electron spin and the magnetic moment. Gao~\cite{GaoJOPCO22} showed that a confined electron has a stable circulating current density, but the wave packet of a free electron discussed by Ohanian is not stable and de-coherences quickly because the wavepacket is not constructed by a pure state.

In this paper we continue the discussion of wave spin. To first show that spin is an embedded property of the electron wave, we derive explicit expressions for the momentum and current densities of an electron in an eigenstate wavefunction $\Psi$ of the Dirac equation, where $i\hbar\frac{\partial}{\partial t}\Psi=\mathcal{E}\Psi$ and $\mathcal{E}$ is the eigen energy. These equations are as follows: 
\begin{eqnarray} \label{GJ} 
&&\pmb{j}=\frac{ec^2}{\mathcal{E}}\left\{\pmb{\nabla}\times\left(\Psi^{\dagger}\frac{\hbar}{2}\pmb{\Sigma}\Psi\right)+i\frac{\hbar}{2}\left[\left(\pmb{\nabla}\Psi^{\dagger}\right)\Psi-\Psi^{\dagger}\left(\pmb{\nabla}\Psi\right)\right]\right\},\nonumber\\ 
&&\pmb{G}=\left\{\frac{1}{2}\pmb{\nabla}\times\left(\Psi^{\dagger}\frac{\hbar}{2}\pmb{\Sigma}\Psi\right)+i\frac{\hbar}{2}\left[\left(\pmb{\nabla}\Psi^{\dagger}\right)\Psi-\Psi^{\dagger}\left(\pmb{\nabla}\Psi\right)\right]\right\}.\nonumber\\ 
\end{eqnarray}
These equations show that both the current density $\pmb{j}$ and the momentum density $\pmb{G}$ contain the spin term with the spin operator $\frac{\hbar}{2}\pmb{\Sigma}$ and the translation term with the momentum operator $-i\hbar\pmb{\nabla}$. It is worth noting that these expressions are derived from different origins. The momentum density is derived from a symmetrical energy-momentum tensor known as the Belinfante-Rosenfeld tensor~\cite{Rosenfeld40}. This symmetrical tensor is constructed to serve as a source for the gravitational field as required by the general relativity. The current density, on the other hand, is derived from the definition $\pmb{j}(\pmb{x})=ec\Psi^{\dagger}(\pmb{x}) \pmb{\alpha}\Psi(\pmb{x})$ using the Gordon decomposition, where $\pmb{\alpha}$ is the $\alpha-$matrix in the Dirac equation. This definition ensures the conservation of charge for the Dirac electron. The momentum and current densities represent the mechanical and electrical nature of the wave spin, and have the same spin and translation terms, except for the factor $\frac{1}{2}$ in the momentum term, which gives the gyromagnetic ratio $g=2$ for the Dirac field.
 
We intend to direct our attention especially to the current density, because the wave spin manifests itself by the current density through the interaction with the electromagnetic field
\begin{equation}
jA = \rho\phi + \pmb{j}\cdot\pmb{A}.
\end{equation}
The above equation accounts for the full electromagnetic interactions in both classical and quantum electrodynamics. In this work, we will show that the current-field interaction $jA$ not only recovers the conventional spin-field interaction, but also reveals other geometric and topological properties that are missing in the particle spin picture, in particular for electrons in excited states.

\section{\label{sec:Excited}Wave spin in excited states}
To investigate the wave spin in the excited states of a confined electron, we first seek the exact solution of the Dirac equation 
\begin{equation}\label{Dirac}
i\hbar \frac{\partial }{\partial t}\Psi (\pmb{r},t)=\left[c\pmb{\alpha} \cdot (-i\hbar \pmb{\nabla })+\gamma ^0 \text{mc}^2+U(\pmb{r})\right]\Psi (\pmb{r},t),
\end{equation}
in a two-dimensional quantum well
\begin{equation}\label{potential}
U(\pmb{r})=U(x,y)=\begin{cases}
0, -L_{x}<x<L_{x},-L_{y}<y<L_{y} \\
\infty, \text{elsewhere.}  \\
\end{cases}
\end{equation}

The four-component spinor wavefunction is expressed by the separation of the temporal and $z$-coordinate variables 
\begin{equation}\label{Psi}
\Psi (\pmb{r},t)=N e^{-i \mathcal{E}t/\hbar} e^{iP_{z}z/\hbar}\left(
\begin{array}{c}
\mu_A(x,y) \\
\mu_B(x,y) 
\end{array}
\right),
\end{equation}
where $\mathcal{E}$ is the energy, $P_{z}$ is the momentum along $z$-direction, and $\mu_A(x,y)$ and $\mu_B(x,y)$ are two-component spinor wavefunctions.

We now plug Eq.~\ref{Psi} into the Dirac equation Eq.~\ref{Dirac} and set $P_{z}=0$ to obtain the coupled equations
\begin{eqnarray}\label{muAB}
\left(\mathcal{E}-mc^{2}\right)\mu_{A}(x,y)&=&-i\hbar c\left(\sigma _x\frac{\partial }{\partial x}+\sigma _y\frac{\partial }{\partial y}\right)\mu_{B}(x,y); \nonumber \\
\left(\mathcal{E}+mc^{2}\right)\mu_{B}(x,y)&=&-i\hbar c\left(\sigma _x\frac{\partial }{\partial x}+\sigma _y\frac{\partial }{\partial y}\right)\mu_{A}(x,y),\nonumber \\
\end{eqnarray}
where $\sigma _x$ and $\sigma _y$ are the Pauli matrixes. 

Eqs.~\ref{muAB} is combined to obtain a second-order differential equation for $\mu_{A}(x,y)$ 
\begin{equation}\label{muA2}
\left(\mathcal{E}^2-m^2 c^4\right)\mu_A(x,y)=-\hbar ^{2}c^{2}\left(\frac{\partial ^2}{\partial x^2}+\frac{\partial ^2}{\partial y^2}\right)\mu_A(x,y), 
\end{equation}
whose eigen solution for the spin-up electron is found 
\begin{equation} \label{muA}
\mu_{A}(x,y)=\sin [k_{x}(x+L_x)] \sin [k_{y}(y+L_y)]  \left(
\begin{array}{c}
1 \\
0
\end{array} \right),
\end{equation}
where $k_x=\frac{\pi  n_x}{2 L_x};k_y=\frac{\pi  n_y}{2 L_y}$ are the wave vectors and $n_x$, $n_y=1,2,3...$ are the quantum numbers of the eigen states. The eigen energy 
\begin{equation} \label{eigenE}
\mathcal{E}=mc^2 \sqrt{1+\eta^2}
\end{equation}
is quantized by $n_x$ and $n_y$ according to the expression of a dimensionless geometric factor
\begin{equation} \label{eta}
\eta =\sqrt{n_x^2\left(\frac{\lambda _c }{4 L_x}\right)^2+n_y^2\left(\frac{\lambda _c }{4 L_y}\right)^2}
\end{equation}
that measures the dimensions of the quantum well $(L_x,L_y) $ against the Compton wavelength $\lambda_c=\frac{\hbar}{mc}$ at different states $(n_x,n_y)$.

The wavefunction $\mu_{B}(x,y)$ is subsequently derived via Eq.~\ref{muAB} and the complete four-component spinor wavefunction is then obtained
\begin{widetext}
\begin{equation} \label{Fullwavefunction}
\Psi (\pmb{r},t)=N e^{-i \mathcal{E}t/\hbar} \left(
\begin{array}{c}
 \sin [k_{x}(x+L_x)] \sin [k_{y}(y+L_y)]  \\
 0 \\
 0 \\
 -i \frac{\eta _x}{1+\sqrt{\eta ^2+1}}\cos [k_{x}(x+L_x)] \sin [k_{y}(y+L_y)] +\frac{\eta _y}{1+\sqrt{\eta ^2+1}}\sin [k_{x}(x+L_x)] \cos [k_{y}(y+L_y)] \\
\end{array}
\right).
\end{equation}
\end{widetext}
where $\eta _x=\frac{\hbar  k_x}{mc}, \eta _y=\frac{\hbar  k_y}{mc}  $ are dimensionless factors along $x$ and $y$ directions. The normalization factor of the wavefunction is found 
\begin{equation}
N=\sqrt{\frac{1+\sqrt{1+\eta^2}}{\sqrt{1+\eta^2}}}.
\end{equation}

The wavefunction in Eq.~\ref{Fullwavefunction} is used to calculate the stable charge density $ \rho(x,y)=e\Psi ^{\dagger } (x) \Psi (x) $ inside the quantum well
\begin{eqnarray} \label{ChargeDensity}
&&\rho(x,y)=e N^2 \sin^{2} [k_{x}(x+L_x)] \sin^{2} [k_{y}(y+L_y)]  \nonumber \\
&& +e N^2 \frac{\eta _x^2}{(1+\sqrt{1+\eta ^2})^2} \cos^{2} [k_{x}(x+L_x)] \sin^{2} [k_{y}(y+L_y)]\nonumber \\
&& +e N^2\frac{\eta _y^2}{(1+\sqrt{1+\eta ^2})^2}\sin^{2} [k_{x}(x+L_x)] \cos^{2} [k_{y}(y+L_y)]. \nonumber \\
\end{eqnarray}
and the corresponding current density
\begin{eqnarray} \label{jxy}
j_x&=& ec \frac{2\eta _y}{\sqrt{1+\eta ^2}}\sin^2 [k_{x}(x+L_x)] \sin  [2k_{y}(y+L_y)], \nonumber \\
j_y&=&-ec \frac{2\eta _x}{\sqrt{1+\eta ^2}}\sin^2  [k_{y}(y+L_y)] \sin [2k_{x}(x+L_x)]. \nonumber \\
\end{eqnarray}

\begin{figure} 
\includegraphics[width=0.48\textwidth]{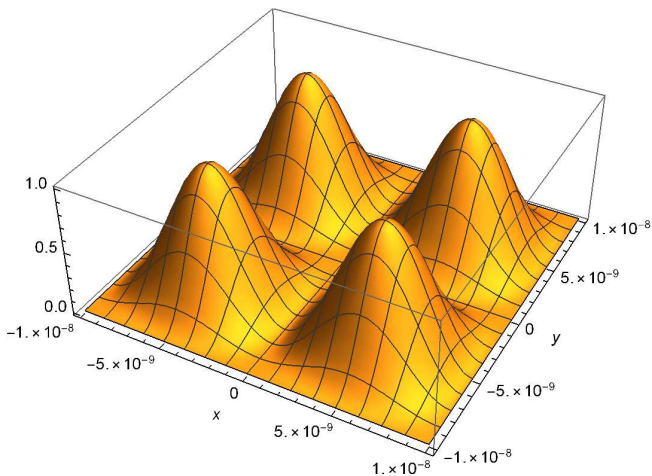}%
\caption{\label{fig:j03D}Charge distribution for the excited state $n_x=2$ and $n_y=2$ for a Dirac electron in a quantum well of $L_x=10~\text{nm}$ and $L_y=10~\text{nm}$. The z-axis represents the charge density of the relative unit.}
\end{figure}

\begin{figure}
\includegraphics[width=0.48\textwidth]{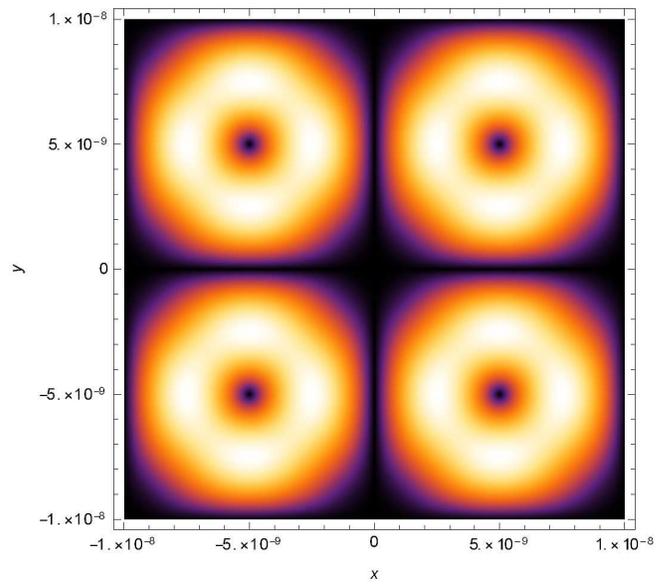}%
\caption{\label{fig:jDensity}Density plot of the current density circulates for the excited state $n_x=2$ and $n_y=2$ for a Dirac electron in a quantum well of $L_x=10~\text{nm}$ and $L_y=10~\text{nm}$.}
\end{figure}

\begin{figure}
\includegraphics[width=0.48\textwidth]{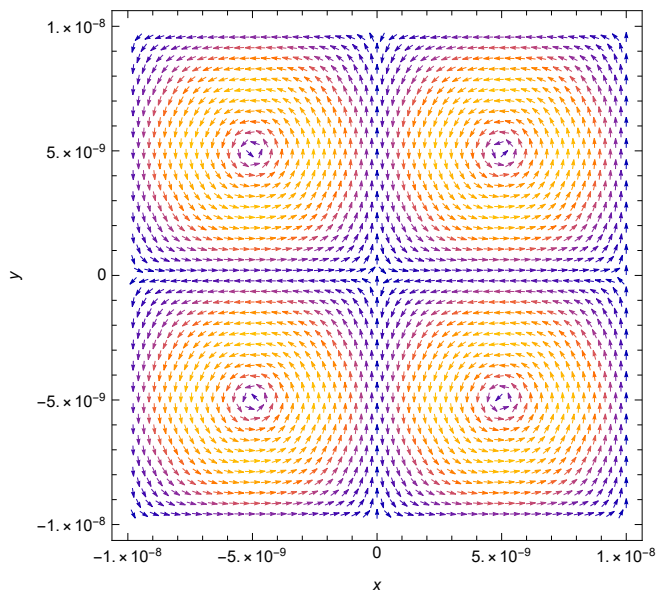}%
\caption{\label{fig:jvector}Vector plot of the current density for the excited state $(n_x=2, n_y=2)$ for a Dirac electron in a quantum well of $L_x=10~\text{nm}$ and $L_y=10~\text{nm}$.}
\end{figure}

Both the charge density and the current density exhibit properties of a standing wave, characterized by the quantum numbers $(n_x,n_y)$. As an example, we choose an excited state $(n_x=2,n_y=2)$ of a quantum well $(L_x=10~\text{nm}, L_y=10~\text{nm})$. Fig.~\ref{fig:j03D} is the density plot of the charge expected for the behavior of an electron cloud. Fig.~\ref{fig:jDensity} is the density plot of the current density of the same electron, showing multiple vortices around the peaks of the electron cloud. The wave spin picture is further illustrated by the vector plot of the current in Fig.~\ref{fig:jvector}, which shows the circulation of each vortex in the same direction. It is clear that the wave spin in the excited state is distributed among multiple vortices that are holographic to each other. This topology suggests that each wave spin vortex represents a part or fraction of the total spin that can be studied and observed by interaction with an external field, as we will elaborate later.   

In Fig.~\ref{fig:jvector} it can also be seen that the current flows continuously along the edge of the quantum well. A similar behavior of the edge current is observed and studied in the quantum Hall effect~\cite{Klitzing2020} when a strong magnetic field is applied to topological insulator materials. The intrinsic edge current shown here suggests that topological spin effects can be observed even without the presence of any magnetic field, internally and externally.  

\begin{figure}
\includegraphics[width=0.45\textwidth]{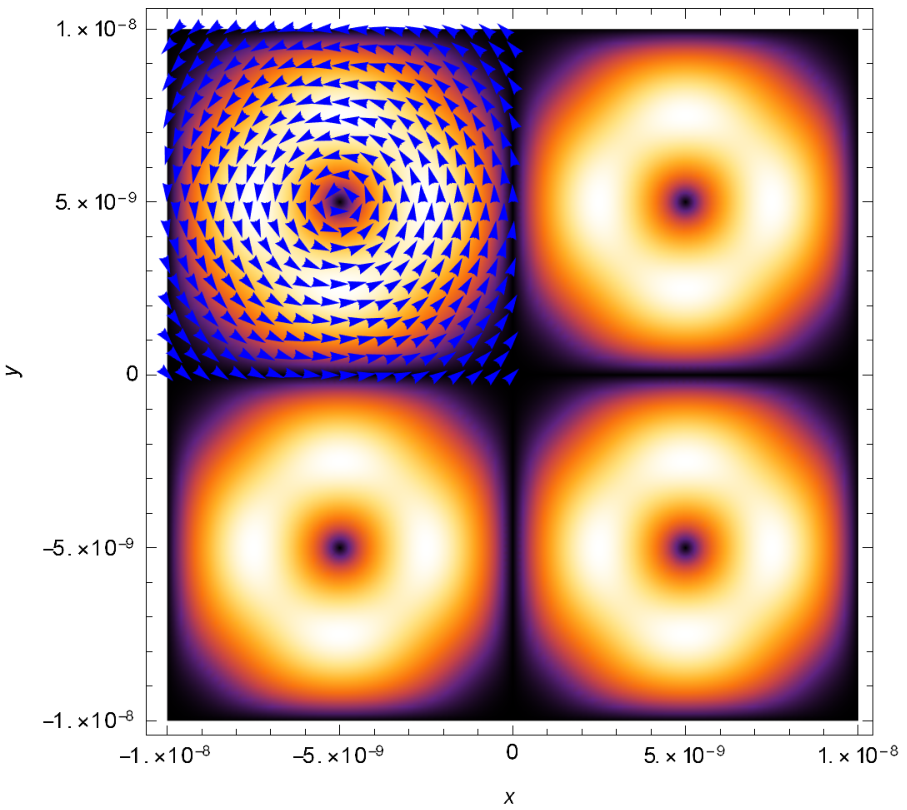}\\
\includegraphics[width=0.45\textwidth]{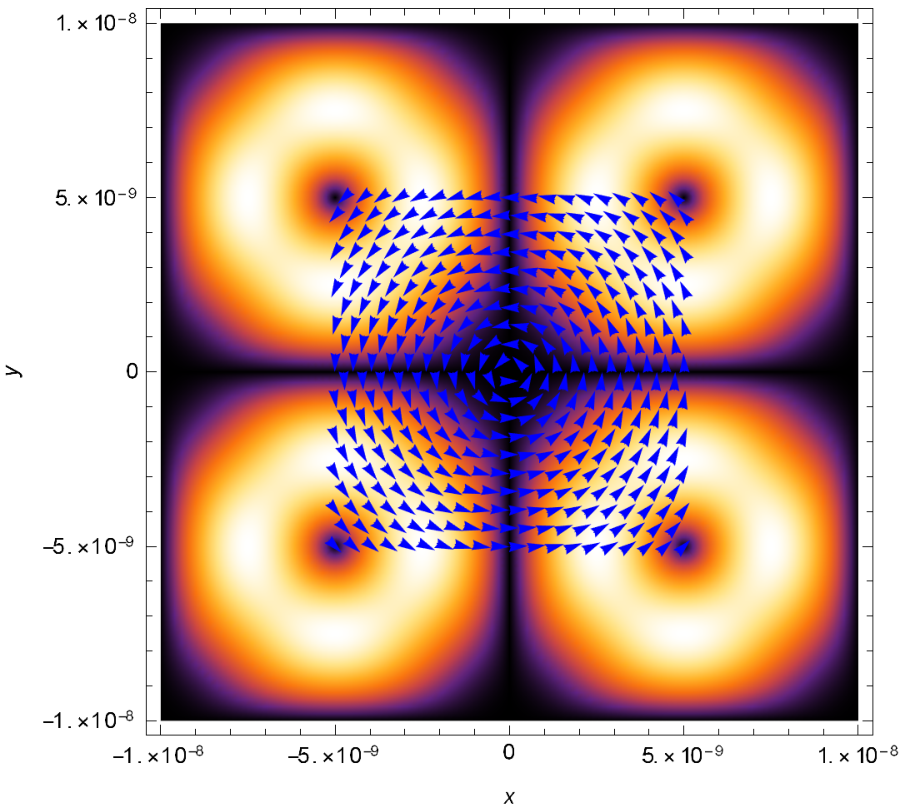}%
\caption{\label{fig:jA12} The upper figure shows the magnetic potential (blue) of Eq.~\ref{Aab} overlapping with the upper left current vortex (red) at $ a=-L_x/2, b=L_y/2$, producing a quarter spin-field coupling. The lower figure shows that the magnetic potential (blue) of Eq.~\ref{Aab} lies in the middle of the quantum well at $ a=0, b=0$, creating zero spin-field coupling.}
\end{figure}

\section{\label{sec:Diss}Topological current-field interaction}
The interaction of a particle spin with an external field is expressed by $\frac{e}{m} \frac{\hbar }{2}\pmb{\Sigma}\cdot \pmb{B} $, which excludes all geometrical and topological effects since $\pmb{\Sigma}$ depicts a dimensionless point. However, since the wave spin is encoded in the current density, its topological property should be transferred to the interaction 
\begin{equation}\label{jA2}
\pmb{j}\cdot \pmb{A}=\Psi ^{\dagger }(\pmb{r},t)ec\pmb{\alpha}\cdot \pmb{A}(\pmb{r})\Psi (\pmb{r},t),
\end{equation}
where the vector potential $ \pmb{A}(\pmb{r}) $ may itself be a vortex field  
\begin{equation}\label{A}
\pmb{A}(x,y)=\frac{B}{2}(-y,x,0)
\end{equation}
to represent a uniform magnetic field along the $z$ direction by the definition $\pmb{\nabla }\times \pmb{A}(x,y)=(0,0,B) $. In this case, this interaction shall be characterized as the vortex-vortex interaction. When the vector potential $ \pmb{A}(\pmb{r}) $ is larger in size than the electron wave, which is usually the case, the interaction of Eq.~\ref{jA2} can be evaluated by integration over the entire quantum well, yielding the following
\begin{eqnarray} \label{E1}
\mathcal{E}^{(1)}&=&\frac{1}{2 L_x}\frac{1}{2 L_y}\int _{-L_y}^{L_y}\int _{-L_x}^{L_x}\pmb{j}\cdot \pmb{A}(x,y)dxdy \nonumber \\
&=&\frac{e \hbar B}{2 m\sqrt{1+\eta ^2} }=\frac{\mu _B B}{\sqrt{1+\eta ^2} },
\end{eqnarray}
where $ \mathcal{E}^{(1)} $ is the first-order energy shift for a weak field and $\mu _B=\frac{e \hbar}{2 m} $ is the Bohr magneton. Thus, the difference between the spin-up and spin-down energy shifts is 
\begin{equation}\label{Zeeman}
\Delta \mathcal{E}=2\mathcal{E}^{(1)}=\frac{2 \mu _B B}{\sqrt{1+\eta ^2}},
\end{equation}
which recovers the expression for the anomalous Zeeman splitting $ 2 \mu _B B $ modified by additional geometric factors and quantum numbers via Eq.~\ref{eta}. This indicates a finer structure of the anomalous Zeeman effect. Since in the wave spin picture the spin state and the spatial state are coupled in the spinor expression, each excited state corresponds to a unique spin state of a unique topology to interact with an external field. 

To show the topological footprints of the vortex-vortex interaction described by Eq.~\ref{jA2}, we postulate a vector potential smaller in size than the electron wave,
\begin{equation}\label{Aab}
\tilde{\pmb{A}}  (\pmb{r})=\begin{cases}
\frac{B}{2}(-y+b,x-a,0), \begin{array}{c}
-L_{x}/2<x-a<L_{x}/2,\\
-L_{y}/2<y-b<L_{y}/2 \\
\end{array} \\ \\
0, \text{    elsewhere.}  
\end{cases}
\end{equation}
to represent a field enclosed in a square that is only a quarter of the quantum well with center $(a, b)$. The vector potential of Eq.~\ref{Aab} produces the same uniform magnetic field within the enclosed region. Fig.~\ref{fig:jA12} shows the vector plot of such a field superimposed on the electron current distribution. We now perform the same integration over the quantum well as in Eq.~\ref{E1}. We find that the current-field interaction depends on the relative position of the vortices. The energy shift for $(n_x=2, n_y=2)$ is now
\begin{eqnarray} \label{E1ab}
\mathcal{E}^{(1)}_{a,b}&=&\frac{1}{2 L_x}\frac{1}{2 L_y}\int _{-L_y}^{L_y}\int _{-L_x}^{L_x}\pmb{j}\cdot \tilde{\pmb{A}} (x,y)dxdy \nonumber \\
&=&\begin{cases}
\frac{1}{4}\frac{\mu _B B}{\sqrt{1+\eta ^2} }, a=\pm L_x/2, b=\pm L_y/2 \\
~~~~0, ~~~~~a=0,b=0,\\
\end{cases}
\end{eqnarray}
which is only $ \frac{1}{4} $ of the spin and magnetic field interaction in Eq.~\ref{E1} when the field overlaps with one of the current vortices. The fractional spin effect is due to the partial participation of the wave spin represented by the current density. A special situation arises when the field lies at the center of the quantum well. Then a zero interaction is observed, because equal fractions of the current flowing in different directions are exactly cancelled out.

The above results are certainly not to be expected from the particle spin picture and should be considered as properties of the wave spin alone. It is conceivable that we can image the entire current density profile by scanning the field against the current distribution. It is now possible to generate vortex optical fields~\cite{Fatkhiev2021} that can be focused on a smaller region than the quantum well. Then the interaction of wave spin and orbital angular momentum can be studied via the vortex-vortex interaction on a controllable scale. All of these discussions and developments mean that we can study and manipulate partial spins to gain knowledge and control over the entire wave spin due to the holographic nature of the multi-vortex current density.

\section{\label{sec:Diss}Discussions and Conclusions}
In summary: 
\begin{enumerate}
\item We argue that spin is not an abstract two-valued property of the electron particle but a property of the electron wave that can be fully described by its momentum and current densities.
\item We show that in the excited state, the current density of an electron forms multiple vortices in a magnetic field-free quantum well. The topology of these vortices depends on the quantum numbers of the states and is holographic in nature.
\item We investigate the anomalous Zeeman effect of the wave spin and show that the anomalous Zeeman splitting contains finer structures than those of the conventional particle spin picture.
\item We show that the geometrical and topological properties of the wave spin can be observed and studied through its interaction with an electromagnetic field. By studying the interaction with a field smaller in size than the wave spin itself, we show that fractional and even zero spin effects can be observed. 
\end{enumerate}

Apparently, the discussion of wave spin could have implications in many areas, and each area requires in-depth study. Here we offer some preliminary discussions.

\begin{enumerate}
\item In the field of quantum technology, the electron spin has been proposed as a candidate for use as a quantum bit (qubit), which is a superposition of the spin-up and spin-down states $\alpha \left(\begin{array}{c} 1 \\ 0\end{array}\right)+\beta \left(\begin{array}{c} 0 \\ 1\end{array}\right)$, where $\alpha$ and $\beta$ are complex numbers. In the wave spin picture, the spin states $\left(\begin{array}{c} 1 \\ 0\end{array}\right)$ and $\left(\begin{array}{c} 0 \\ 1\end{array}\right)$ are replaced by the spinor wave functions, as in Eq.~\ref{Fullwavefunction}, where the spatial wave function serves as the spinor component. Therefore, we have argued that the spin cannot be completely isolated from the physical environment in which it resides. Any change of the boundary conditions alters the wave functions, and hence the spin states. We further argue that the spin cannot be completely isolated from the Hilbert space of the electron either. Any transition to or from other spatial states alters the original spin state, leading to decoherence of the prepared qubit and loss of the quantum information. This means that additional protection and correction mechanisms need to be implemented to protect and preserve the spin qubit.
\item The multi-vortex wave spin topology and its partial interaction with electromagnetic fields show that the spin-field interaction is not dimensionless, suggesting novel schemes for parallel information processing using spin. Each vortex of the current is a holographic part of the entire spin and can interact simultaneously with multiple electromagnetic fields, potentially enabling parallel computing. Building such parallel computers shall benefit from the advancement of both spintronics and optics. 
\item It is conceivable that the holographic spin interactions could already exist in nature. It is suggested that the electron spin could play a role in bio-homochirality~\cite{Wangwei}, which makes the wave spin perspective interesting for this biological topic~\cite{RiveraIslas2004} and the general spin effects in molecules. The wave spin could interact holographically with all atoms within the molecule via the shared electron cloud, leaving a coherent spin footprint on the entire molecular structure.  
\end{enumerate} 

\section{\label{sec:Acknowledgement}Acknowledgement}
The authors would like to thank Jora L. Gao for her careful review of the manuscript. The authors would also like to thank Jane Y. Gao and W. Wang for stimulating discussions on the role of spin in biology.

\bibliography{Spin}% Produces the bibliography via BibTeX.

\end{document}